# A Virtual World Model to Enhance Tourism Destination Accessibility in Developing Countries

Arunasalam Sambhanthan[1], Alice Good[2]

School of Computing, University of Portsmouth, UK

arunsambhanthan@gmail.com[1], alice.good@port.ac.uk[2]

## ABSTRACT

The problem of destination accessibility is a vital concern in the sustainable tourism development in the emerging regions due to the increasing numbers of tourism business growth in the recent times. Tourism is one of the potential foreign exchange earning sectors, which place sustainability as one of the main success metrics for benchmarking the industry's overall development. On the other hand, there are several destinations, which are inaccessible to tourists due to several reasons. Underutilization of potential destinations in both pre purchase and consumption stages is a strategic disadvantage for emerging countries on leading their tourism industry towards sustainability. A virtual world model to increase the destination accessibility of tourism products has been proposed. The model has to be designed with visual and auditory experience to tourists. The model is expected to enhance the accessibility of destinations for users of different categories. Elderly users, users with panic disorders, users with mobility impairments also will be able to enjoy traveling experience just same as other, through the proposed model.

## INTRODUCTION

The question of sustainability in tourism development is a contemporary concern. In other words, developing tourism in a manner which does not compromise the preservation of tourism resources for the future generations is a critical concern. The present global credit crunch has further increased the degree of challenge involved in ensuring sustainable tourism development. The emerging countries, being vulnerable to the above dilemma need to adapt innovative methods to ensure sustainable development of their tourism business. Arguably, tourism is one of the potential foreign exchange earning sectors, which place sustainability as one of the main success metrics for benchmarking the industry's overall development. On the other hand, there are several destinations, which are inaccessible to tourists due to several reasons. Underutilization of potential destinations in both pre purchase and consumption stages is a strategic disadvantage for emerging countries on leading their tourism industry towards sustainability.

Firstly, the quality of product information largely relies on the degree of experience added with the information. In particular, the travel experience delivered online in terms of visual and auditory forms play a vital role in influencing the purchase decision of tourists. This entails the need for bridging the gap between product information and the exact experience. Improved quality of information in the pre purchase stage will increase the revenue by positively influencing the purchase behavior of consumers. Also, it will enhance the post consumption experience by reducing the mismatch between expectations and exact features. Secondly, tourists with impairments and elderly tourists get deprived from consuming the tourism product, as same as the normal tourists. This deems a need for bringing the destinations to their doorstep, with all its real world experiences to enhance tourism development in the consumption stage. Thirdly, there are inherently endangered tourism destinations, which cannot be accessed by human beings due to complex geographical location. Enablement of tourists to get access to these destinations is another open question, which could result in the development of new tourism products. Hence, a solution to enhance the accessibility of tourism destinations in all these three means is an evolving contemporary need for successful tourism development. This paper, proposes a virtual world model to bridge the above gap, along with a number of distinct value additions to tourism development. This means the proposed model could enhance the accessibility of tourism destinations to people with impairments through delivering a tourism experience in the virtual world. This will be achieved through making these countries accessible in terms of destination for the impaired person as well as ensuring the interface is accessible to all users.

## Literature Review

Interactions in Virtual Worlds are essentially about avatar interactions. Avatar interactions are quite different and unique compared to the Web 2.0 based interactions which are solely between the interface and the user. According to Davis *et al.* (2005) "An avatar is defined as a user created digital representation that symbolizes the user's presence in a metaverse".

Some companies are using avatar based innovations as a way forward to new product development. To clarify, the avatar-based approach adapted in the above context opens up avenues for innovative product design and content creation in the virtual world (Kohler, *et al.*, 2009). They argue that the avatar based innovations present an opportunity for companies to engage with customers in new and innovative ways during an interactive new product development process. The user generated nature of the virtual platform could further enrich innovation efforts. Research on how the companies could attract appropriate customers, and which incentives the firms need to implement in order to promote and leverage valuable customer contributions, could provide some open avenues for further exploration.

Additionally, to which extent an avatar is engaging in deceptive behavior or what measures the companies could take to guarantee the quality in contributions - are some potential areas for further research. Furthermore, the nature of Virtual Worlds calls for a re-examination of various issues such as avatar motivation to engage in co-innovation activities and the degree of interaction efficiency among avatars. The question arises as to what are the mechanisms for supporting and facilitating collaborative innovation in Virtual Worlds? According to Kohler *et al.* (2009), further research is also warranted to compare traditional web based methods with avatar based efforts, to shed light on the question of when to best employ which technology. It is also important to study their use for diverse new product development tasks due to the market forces, niche opportunities, technological development, and so on. However, this research does not appear to focus on the interaction related issues pertaining to the avatar based innovations. Particularly, the psychological concerns pertaining to the avatar based interactions will pose challenges to the effectiveness of interactions compared to real world interactions.

Another important use of Virtual Worlds is the ability to collaborate in real time through synchronous communication medium. Virtual Worlds allow globally dispersed teams to collaborate and work in a virtual environment, where the avatar based interaction happens as part of the collaboration (Davis et al., 2009).

A recent study explores on the potential of Virtual Worlds in enriching innovation and collaboration in Information Systems research, development and commercialization (Dreher et al., 2011). The authors argue that the Virtual Worlds - by their very structure provide a powerful context for innovation and collaboration. Their paper concludes stating that there is great potential inherent in the use of 3D Digital Ecosystems for Information Systems Technology research, development, and commercialization. Such developments will keep pace with the digital-native culture of younger generations and have the potential to innovatively revolutionize our social systems relating to governance, education, commerce, and social interaction. Digital Ecosystems, 3D Virtual Worlds in particular, are set to lead the charge in our modern culture of accelerating innovation (Dreher et al., 2011).

Another recent study by Eklund *et al.* (2009) reports, a Virtual Museum of the Pacific - implemented as Web 2.0 application that experiments with information and knowledge acquisition for a digital collection of museum artifacts from the Australian Museum. Hence the mission of this paper is to evaluate the emerging

Virtual World models with regard to the socio-political, technological and ethical aspects and to evaluate the degree of contributions made by these emerging models to the body of knowledge.

Messinger *et al*. (2008) discussed the typology for Virtual Communities, and the historical developments of Virtual Worlds research, clearly outlining the development of the gaming industry as well as the social networking industry. The field of Virtual Worlds is a unique blending of both these industries over a few decades.

**Virtual Reality in Tourism Development**

Steuer (1993) argues that 'presence' and 'telepresence' as the underlying conceptual elements of virtual reality. Particularly the sense of being in an environment is the main requirement, which could be generated by natural or mediated means. Cheong (1995) defines virtual reality as "a computer mediated sensory experience that serves to facilitate access into dimensions that differ from our own". However, the above definition does not have an operational level focus, but more of an abstract level statement. On the other hand, there are certain constraints involved in creating a computer mediated sensory experience. In particular, there are certain technological constraints involved in developing a system which could generate experience to satisfy all five senses. Although it is possible through the state of the art virtual reality studies, it is a challenge to implement such a system in emerging countries due to the infrastructural constraints involved. Particularly, the organizations in the emerging countries would not be able to afford huge amount of money on a virtual reality system, while there are several other priorities to be addressed while making investment decisions in tourism development. Pizam (2009) indicates that the global financial crisis has immensely impacted the global travel industry. Hence, there are certain financial constraints involved in producing a complete virtual experience of five senses through a system in the context of emerging regions. Therefore Virtual Reality is defined as "a computer mediated sensory experience which facilitate access to visual and auditory dimensions of a travel destination" in the course of developing this model. The definition provides a contextual relevance to the requirements deemed by the travel related endeavors of emerging regions.

A recent study claims game animation technology as the possible next innovation in tourism (Tjostheim et al, 2005). But the above study evaluated the context of an entire 'virtual tour' facilitated through WWW. Also this research was tested with retail business - a non travel industry. However, the user testing done by Tjostheim et al (2005, p 9) with regard to the game animation technology utilization indicates that a typical home system will be enough to facilitate the above technology. Also the study reports a high technical compatibility of the particular technology. However, contextualizing the findings to tourism domain is an obvious challenge in front of practitioners. Above findings, in light of the current period of financial downtrodden urges the Sri Lankan hoteliers to go for middle approach between a complete virtual tour and passive web presence. Consequently, a part application of simulation technology allowing a small glimpse of tours experience to be received through the website will better facilitate excellence in web tourism promotion.

**Virtual Museum of the Pacific**

The virtual museum of the pacific is another interesting case to be evaluated. The authors describe the design process adapted for building the digital museum project hosted at the University of Wollongong. The paper describes the Virtual Museum of the pacific as a digital ecosystem in which objects of a digital collection of museum artifacts are derived from facets of the physical artifacts held in the Australian Museum's pacific collection. The virtual museum of the pacific allows several diverse search methods: attribute search based on a control vocabulary, search via query refinement and query by example. Further to this the system also provides a number of management interfaces that enable content to be added and tagged, the control vocabulary to be extended, user perceptions to be defined and narratives added via wiki.

In evaluating the typological aspect of this case, the purpose is defined to be as providing museum access to the large audience through the virtual museum access to large audience through the virtual museum interface. The place of interaction is set to be completely Virtual through the interface designed as part of this project. Platform of this case is through a synchronous communication mechanism facilitated through a query refinement approach. The population or pattern of interaction in this case is defined as the large target audience focuses through the museum. Finally, the project model would be the subscription fee or payments such as endowments provided by the user as part of / as a result of their interaction.

**Virtual Hats – A Role Playing Activity**

Role playing in Virtual Worlds have a tremendous potential for allowing students to have effective learning endeavors due to the synchronous communication. Also, obviously Virtual Worlds have the potential to facilitate students engage in learning activities which are not possible in the real world. In this article, a project that involved pre-service teachers carrying out role-plays based on de Bono's *Six*

*Thinking Hats* framework is presented. A pilot study was carried out over two years with on-campus students, who performed the role-plays both in a real-life, physical setting and within the virtual world of *Second Life*. In overall the study presented argues that real world setting could be simulated or replicated in the virtual words. The results of the pilot study suggest that students have a preference for real-life, face-to-face learning activities; however, the participants in the study were on-campus students, who, unlike those who are studying at a distance, actually have the privilege of access to this mode of learning.

**Otago Virtual Hospital**

Otago Virtual Hospital model is designed to formatively assess dispositional behaviors in scenario based in the Virtual Worlds. The framework was devised for use with medical students playing the roles of junior doctors as they solve open ended clinical cases within an environment called the Otago Virtual Hospital. In doing this, the authors designed a conceptual framework in which medical students are retrospectively assessed based on the number of times they either seized or missed an opportunity to engage in a particular dispositional behavior. In this article, the authors have presented an empirical illustration of our conceptual framework.

### Virtual World Model for Tourism Development

The model aims to enhance the tourism development of specific regions by increasing the destination accessibility, through which the economic sustainability could be dramatically improved. Specifically, the proposed model will enhance the accessibility of tourism destinations to tourists with mobility impairments as well elderly tourists. The experiential nature of VR provides more sensory and rich information about the destinations to tourists (Guttentag, 2010). Hence, the purchase decision of tourists could be immensely influenced through this model. This will lead to a major economic gain through increasing the number of visits to the destinations through virtual world. In addition to this, the model could be used as an advanced promotional tool for tourism through which a prospective tourist could gain a more advanced understanding of the features of specific tourism destinations. Also, researches argue that an e-tourism experience would immensely helps tourists with panic disorder to get rid of unnecessary dismay during the exact visit (Newman, 2008). Furthermore, the model could possibly used to substitute the travel experience of endangered destinations, which are inaccessible to even normal tourists. In conjunction, the model could also be used for environmental gains related to tourism development. Especially, the eco-efficiency of tourism destinations could be indirectly enhanced through increasing the virtual visits instead of trouping horde of people across delicate habitats. Particularly the solution could immensely contribute towards the economic development of developing countries, which earns a considerable amount of annual foreign exchange through tourism.

## Further Enhancements

The above model could be utilized for enhancing the overall development of tourism business in emerging regions. Firstly, the model could be utilized for developing virtual theme parks. Development of virtual theme parks has already been identified as a potential benefit of virtual reality in tourism business (Williams & Hobson, 1995). However, it is a challenge for emerging regions to spend huge money on theme parks, that bottleneck could be complemented through developing virtual theme parks. On the other hand, the model could be used as a means of enhancing ecotourism in the regions. Especially, reducing the amount of actual visits to the destinations would reduce the environmental effect and possibly contribute towards the development of ecotourism industry. In addition to this, the model could be used as a means to replicate some endangered destinations which are not accessible to human beings.

## Bibliography

Butler, D. (2012). Second Life Machinima enhancing the learning of Law: Lessons from Successful Endeavors, *Australian Journal of Educational Technology*.

Cheong, R. (1995), The Virtual Threat to Travel and Tourism, Tourism Management, vol. 16, no. 6, pp. 417 – 422.

Davis, A., Khazanchi, D, Murphy, J. Zigurs, I and Owens, D. (2009), Avatars, People and Virtual Worlds: Foundations for Research in Metaverses, *Journal of the Association of Information Systems,* vol. 10, no. 2, pp. 90-117.

Dreher, C., Reiners, T., Dreher, N. and Dreher, H. (2011), 3D Virtual Worlds as Collaborative Communities Enhancing Human Endeavors: Innovative Approaches in e-learning, *Proceedings of 3rd IEEE International Conference on Digital Ecosystems and Technologies*, pp. 138-143.

Dreher, C., Rieners, T., Dreher, N. and Dreher, H. (2011), 3D Virtual Worlds Enriching Innovation and Collaboration in Information Systems Research, Development and Commercialization, *Proceedings of 3rd IEEE International Conference on Digital Ecosystems and Technologies*, pp. 168-173.

Eklund, P., Goodall, P., Wray, T, Bunt, B. and Lawson, A. (2009), Designing the Digital Ecosystem of the Virtual Museum of the Pacific, *Proceedings of 3rd IEEE International Conference on Digital Ecosystems and Technologies*, pp. 377-383.

Gregory, S. and Masters, Y. (2012), Real thinking with Virtual Hats: A role playing activity for pre-service teachers in second life, *Australian Journal of Educational Technology*.

Guttentag, D. A. (2010), Virtual Reality: Applications and Implications for Tourism, Tourism Management, vol. 31 (5), pp. 637 651.

Hendaoui, A., Limayem, M., and Thompson, C.W., (2008), 3D Social Virtual Worlds: Research Issues and Challenges, *IEEE Internet Computing*, pp. 88-92.

Henderson, M., Huang, H., Grant, S. and Henderson, L. (2012). The impact of Chinese Language Lessons in a Virtual World on University Students' Self Efficacy Beliefs, *Australian Journal of Educational Technology*.

Kohler, T. Matzler, K and Fuller, J. (2009), Avatar based Innovation: Using Virtual Worlds for real world Innovation, *Technovation*, 29, pp. 395 – 407.

Messinger, P.R., Stroulia, E, Lyons, K. (2008), Virtual Worlds Research: Past, Present & Future, *Journal of Virtual Worlds Research*, vol.1, no.1, pp. 2-18.

Newman, T., W. (2008), Imaginative Travel: Experiential Aspects of User Interactions with Destination Marketing Websites, Doctoral Thesis, Auckland university of Technology, Auckland, New Zealand, Retrieved on 08th September 2009 from http://aut.researchgateway.ac.nz/handle/10292/665?mode=simple.

Pizam, A. (2009). The Global Financial Crisis and Its Impact on the Hospitality Industry. *International Journal of Hospitality Management*, 28, 301.

Steuer, J. (1993), Defining Virtual Reality: Dimensions Determining Telepresence, Journal of Communication, autumn, 1992, pp. 73 – 93.

Stewart, S. and Davis, D. (2012), On the MUVE or in decline: Reflecting on the Sustainability of the Virtual Birth Centre developed in Second Life, *Australian Journal of Educational Technology*.

Tjostheim, Ingvar, Lous, Joachim (2005). A Game Experience in Every Travel Website?

Wegener, M., McIntyre, T.J., McGrath, D., Savage, C.M., Williamson, M., Developing a Virtual Physics World, Australian Journal of Educational Technology, 28 (3), pp. 504-521.

Williams, P. & Hobson, J.S.P., (1995), Virtual Reality and Tourism: Fact or Fantasy?, Tourism Management, vol. 16, no. 6, pp. 423 – 427.

Wimpenny, K., Savin-Baden, M. Mawer, M., Steils, N. and Tombs, G. (2012), Unpacking frames of reference to inform the design of Virtual World Learning in Higher Education, Australian Journal of Educational Technology, 28 (3), pp. 522-545.

Wood, D. and Willems, J. (2012), Responding to the widening participation agenda through improved access to and within 3D virtual learning environments, *Australian Journal of Educational Technology*.